\documentstyle[amssymb,prb,aps,graphicx]{revtex}

\begin{document}
\draft

\twocolumn[\hsize\textwidth\columnwidth\hsize\csname@twocolumnfalse\endcsname
\title{Observation of a controllable $\pi$-junction in a 3-terminal Josephson device}
\author{Jian Huang$^{1}$, F. Pierre$^{1}$, Tero T. Heikkil\"{a}$^{2}$, Frank K.
Wilhelm$^{3}$, and Norman O. Birge$^{1}$}
\address{$^{1}$ Department of Physics and Astronomy, Michigan State University, East Lansing, MI 48824-1116\\
$^{2}$ Materials Physics Laboratory, Helsinki University of
Technology, FIN-02015 HUT, Finland\\
$^{3}$Sektion Physik and CeNS, LMU, Theresienstr. 37, D-80333
M\"{u}nchen, Germany}
\date{\today}
\maketitle

\begin{abstract}
Recently Baselmans et al. {[Nature, {\bf 397}, 43, (1999)]} showed
that the direction of the supercurrent in a
superconductor/normal/superconductor Josephson junction can be
reversed by applying, perpendicularly to the supercurrent, a
sufficiently large control current between two normal reservoirs.
The novel behavior of their 4-terminal device (called a
controllable $\pi $-junction) arises from the nonequilibrium
electron energy distribution established in the normal wire
between the two superconductors. We have observed a similar
supercurrent reversal in a 3-terminal device, where the control
current passes from a single normal reservoir into the two
superconductors. We show theoretically that this behavior,
although intuitively less obvious, arises from the same
nonequilibrium physics present in the 4-terminal device. Moreover,
we argue that the amplitude of the $\pi $-state critical current
should be at least as large in the 3-terminal device as in a
comparable 4-terminal device.
\end{abstract}

]


When a normal metal is put in contact with one or more
superconductors, the properties of both materials are modified
near the interface. The physical phenomena associated with
superconductor (S)/normal (N) systems, namely the proximity and
Josephson effects, were intensely studied in the 1960's and
70's.\cite{oldproxy} Interest in S/N systems was rekindled in the
1990's due to the ability to fabricate complex structures with
submicrometer dimensions. A new, deeper understanding of the
proximity effect on mesoscopic length scales has
emerged,\cite{proxy,Lambert} concentrating on equilibrium and
linear-response physics.
\begin{figure}[tbp]
\begin{center}
\includegraphics[width=2.5in]{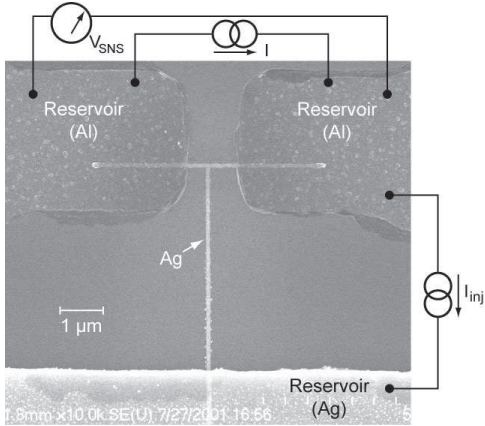}
\end{center}
\caption{Scanning electron microscope picture of the sample, with
schematic drawing of the measurement circuit. The sample consists
of a T-shaped Ag wire with lateral dimensions of 50~nm $\times $
70~nm, connected to two 70~nm thick Al electrodes and one 230~nm
thick Ag electrode.} \label{Fig1}
\end{figure}

Nonequilibrium phenomena in S/N systems are now taking the spotlight.\cite
{nonequilibrium,PierreMAR,Baselmans} A major discovery was made by Baselmans
et al.,\cite{Baselmans} who measured a 4-terminal diffusive metal S/N/S
Josephson device with a cross shape. Two opposing ends of the cross were
connected to S electrodes, while the other two were connected to N
reservoirs between which a control current was passed. Baselmans et al.
found that, at high control current, in samples with the normal reservoirs
sufficiently close together, the sign of the Josephson supercurrent between
the S electrodes reversed direction. The current-phase relationship under
such conditions becomes $I_{s}(\phi )=I_{c}\sin (\phi +\pi ),$ where $I_{c}$%
\ is the (positive) critical supercurrent, rather than the usual Josephson
relationship $I_{s}(\phi )=I_{c}\sin (\phi )$, hence the device is called a $%
\pi $-junction. Such a device has been used to make a controllable $\pi $%
-SQUID.\cite{PiSQUID} The explanation of the nonequilibrium $\pi
$-junction consists of two parts.\cite{PiThy} First, the
supercurrent can be decomposed into an energy-dependent ``spectral
supercurrent'' $j_{E}$, which is an equilibrium property
determined by the sample geometry and resistance as well as the
phase difference $\phi $ between the two S electrodes. $j_{E}$ is
an odd function of energy, and exhibits damped oscillations on an
energy scale comparable to the Thouless energy of the sample,
$E_{th}=\hbar D/L^{2}, $ with $D$ the diffusion constant in the
wire and $L$ the length between the superconductors. Second, the
total supercurrent is determined by the occupation of the
supercurrent-carrying states, given by the antisymmetric part of
the quasiparticle distribution function $f(E)$ in the normal
region of the junction describing the pairs of quasiparticles
($E>E_{F}$) and quasiholes ($E<E_{F}$). Under nonequilibrium
conditions, $f(E)$ can be made to have a staircase shape, with
steps appearing at the voltages of the normal
reservoirs.\cite{PRLrelax} The staircase shape of $f(E)$ excludes
the low-energy contribution of $j_{E}$ from the supercurrent. When
the control voltage approaches the energy where $j_{E}$ changes
sign, the supercurrent
changes its sign relative to the equilibrium situation. In contrast to the $%
\pi $-junction behavior, smearing of the distribution function by electron
heating or raising the sample temperature simply causes the supercurrent to
decrease toward zero without ever changing sign.

The sample shown in Fig. 1 consists of a T-shaped Ag wire, 70~nm wide and
50~nm thick, connected to two S electrodes (70~nm of Al) and one N reservoir
(230~nm of Ag). The distance between S electrodes is 1.1~$\mu$m, while the
distance from the top of the ``T'' to the N reservoir is 4.5~$\mu$m. The
phase coherence length $L_{\phi }$ in similarly prepared Ag wires is several
micrometers at sub-Kelvin temperatures, hence we expect to observe a
substantial Josephson effect between the two S electrodes. The sample was
fabricated using one electron-beam and two optical lithography steps. The
T-shaped Ag wire was fabricated first, followed by the thick Ag reservoir,
and finally the Al electrodes. A gentle ion mill of the exposed ends of the
Ag wire preceded the evaporation of the Al electrodes to enhance the
transparency of the Ag/Al interfaces. The sample was immersed in the mixing
chamber of a dilution refrigerator with filtered electrical leads.

The transport properties of the sample were determined initially
by measuring the $V$ vs. $I$ characteristics between pairs of
electrodes. The $V $-$I$ curve between S electrodes shows the
standard Josephson junction behavior with a critical current of
0.7~$\mu $A at 38~mK. The $V$-$I$ curve between the N electrode
and either S electrode exhibits a change in slope at a current
approximately equal to twice the critical current. This behavior
is due to the superposition of opposite-flowing quasiparticle
current and supercurrent in the dangling arm, as observed recently
by Shaikhaidarov et al.\cite{Shaikhaidarov}. For the sample shown
in Fig.~1, the left and right arms have resistances of
$R_{1}=7.0~\Omega $ and $R_{2}=9.1$~$\Omega $, respectively, while
the base of the T has a resistance of $R_{0}=36$~$\Omega $. From
these values and the sample geometry, we deduce that about half of
the 16.1~$\Omega $ S-S resistance comes from the uncovered part of
the Ag wire, and the other half from the Al/Ag interfaces and part
of the Ag wire extending under the Al electrodes.
\begin{figure}[tbp]
\includegraphics[width=3in]{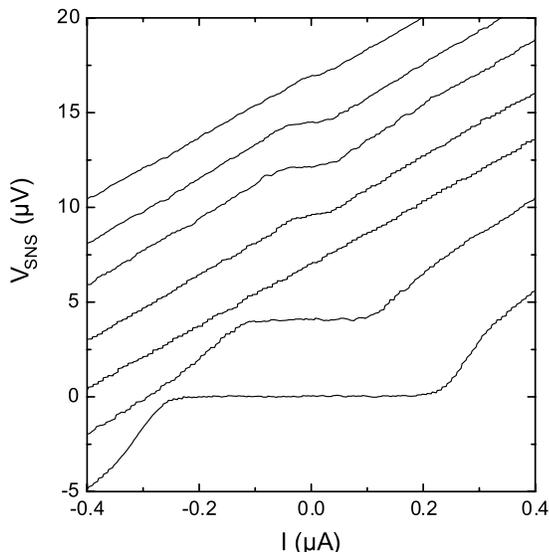}
\caption{A subset of $V_{SNS}$ vs. $I$ curves measured across the
S/N/S Josephson junction, for different values of the current
injected from the normal reservoir. From bottom to top, the
injected currents $I_{{\rm inj}}$ are in $\protect\mu A$: 0.53,
0.70, 1.01, 1.23, 1.89, 2.18, 3.15. The curves are offset for
clarity.} \label{Fig2}
\end{figure}

The measurement circuit for the nonequilibrium injection experiment is shown
schematically in Fig.~1. A dc current $I_{{\rm inj}}$ is injected from the
normal electrode to one of the superconducting electrodes. Simultaneously,
the $V$-$I$ curve between the two superconducting electrodes is measured in
a 4-probe configuration. Figure~2 shows a subset of $V$-$I$ curves for
different values of $I_{{\rm inj}}$, and is the central result of this
paper. The critical current of the S/N/S junction decreases rapidly with
increasing injection current. When $I_{{\rm inj}}=1.0$~$\mu$A, the critical
current is below our measurement threshold. Upon further increase of $I_{%
{\rm inj}}$, the critical current increases again, and finally disappears
when $I_{{\rm inj}}>3$~$\mu$A. In Figure~3, we plot $I_{c}$ vs. $V_{N}$ at
three different temperatures, where $V_{N}=R_{N}I_{{\rm inj}}$ is the
voltage of the normal reservoir with respect to the superconductors, and $%
R_{N}=R_{0}+(R_{1}^{{-1}}+R_{2}^{{-1}})^{{-1}}=40$~$\Omega $. In the figure
we intentionally plot $I_{c}<0$ after it falls to zero, to emphasize that
the junction has entered the ``$\pi $'' state.\cite{signofIc} Our
interpretation of the data is consistent with the assumption that, for fixed
$\phi ,$ $I_{s}$ is a smooth function of $V_{N}$ with a continuous first
derivative. It is also consistent with the experiment of Baselmans et al.,
\cite{Baselmans} who confirmed the existence of the ``$\pi $'' state by
measuring the resistance of the normal wire as a function of the
supercurrent, hence the phase difference $\phi $, between the S electrodes.
At zero supercurrent, their wire resistance exhibits a local minimum in the
usual ``$0$'' state and a local maximum in the ``$\pi $'' state due to the
proximity effect.
\begin{figure}[tbp]
\includegraphics[width=3in]{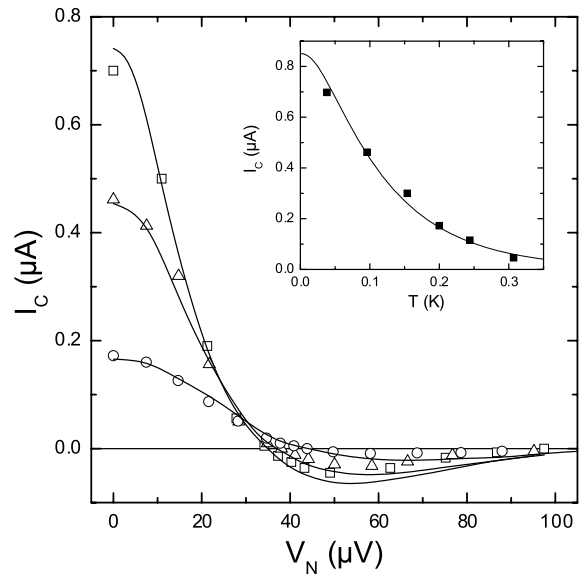}
\caption{Critical current of the Josephson junction vs. voltage of the
normal reservoir at T=38 ($\Box$), 96 ($\triangle$) and 200~mK ($\circ$). \ $%
I_{c}$ is shown as negative for $V_{N} \gtrsim 40\protect\mu V$ to symbolize
the appearance of the $\protect\pi $-junction. Inset: Critical current vs.
temperature at $V_{N}=0$. The lines are the theoretical calculations
discussed in the text.}
\label{Fig3}
\end{figure}

The significant difference between our experiment and that of
Baselmans et al., aside from the reduction from 4 terminals to 3,
is the presence in our sample of a dissipative quasiparticle
current in the sample arms that simultaneously carry the
supercurrent. In the Baselmans experiment, the control voltages of
the two normal reservoirs were set to values $\pm V_{N}$ with
respect to the superconductors, so that the electrical potential
was zero everywhere along the wire connecting the two
superconductors. To compare our experiment with theirs, we must
understand the influence of the dissipative current on the
supercurrent in our sample. We use the quasiclassical formalism in
real time, which was originally developed for nonequilibrium
phenomena in massive superconductors~\cite{RammerSmith} but also
adapted and successfully applied to mesoscopic proximity systems,
as reviewed e.g. in refs.~3 and 14.

For the present paper, we are concerned primarily with the
supercurrent
\begin{equation}
I_{S}=\frac{\sigma _{N}A}{2}\int j_{E}f_{L}(E)dE,
\end{equation}
where $\sigma _{N}$ and $A$ are the conductance and cross-section of the
normal wire, $j_{E}$ is the spectral supercurrent discussed earlier, and $%
f_{L}(E)=f(-E)-f(E)$ is the antisymmetric part of the electron
energy distribution function. With the chemical potential of the
superconductors taken to be zero, the symmetric distribution
function $f_{T}(E)=1-f(E)-f(-E)$ describes charge imbalance, while
$f_{L}(E)$ describes energy or heat in the conduction electron
system.

To calculate the supercurrent, first one must solve the Usadel
equation for the retarded and advanced Green's functions. Those
contain all information about energy-dependent properties of the
sample, including the function $j_{E}$. To find $f_{L}(E)$, one
must then solve the Keldysh component of the Usadel equation,
which takes the form of conservation laws for the spectral charge
and heat currents.\cite{Belzig} When $j_{E}\neq 0$, the two
kinetic equations are coupled, and lead to complicated spatial and
energy dependences of $f_{L}(E)$ and $f_{T}(E)$ in the arms of the
sample between the superconductors. A major simplification occurs
in the arm of the sample connected to the normal reservoir:
$j_{E}=0$ there since the superconducting phase is constant along
that arm. For voltages and temperatures small compared to $\Delta
$ the heat current is zero, \cite{Andreev} hence $f_{L}(E)$ is
constant along that arm and takes on the (equilibrium) value it
has in the N reservoir: $f_{L}^{0}=(1/2)\{\tanh
[(E+eV_{N})/2k_{B}T]+\tanh [(E-eV_{N})/2k_{B}T]\}$. \ Since the
total charge current is conserved along the two sample arms
connecting the superconductors, we can evaluate it anywhere in
those arms. At the central point, the dissipative currents
diverted into the two arms cancel and we can find the supercurrent
from Eq. (1) using the expression for $f_{L}^{0}(E)$ given above,
without integrating the kinetic equations. We need only to
evaluate $j_{E}$ at the central point by solving the equilibrium
Usadel equation for our sample geometry.

As an extension of previous work,\cite{supercurrent} we have
solved the retarded Usadel equation taking into account the
influence of the lead to the normal reservoir and the finite
interface resistances.\cite {DirtyInterface} \ The normal
reservoir induces extra decoherence into the structure, decreasing
the magnitude of the observed supercurrent. We find that the full
gap in the spectral supercurrent\cite{supercurrent} becomes a
pseudogap and that the amplitude of the maximum of $j_{E}$ is
strongly reduced (although the total supercurrent is reduced by
only 20\% at 40~mK). Our fit to the equilibrium data of critical
supercurrent vs. temperature is shown in the inset to Figure~3. To
fit the temperature dependence, the Thouless energy was adjusted
to be $E_{{\rm Th}}=3.5~\mu $eV, which corresponds to a distance
$L=1.7~\mu $m between the superconducting electrodes -- larger
than the actual distance as a result of the silver wire
penetration under the aluminium reservoirs and of the finite
contact resistances. Surprisingly, the magnitude of the calculated
cricital current had to be reduced by a factor 1.7 to match the
experimental data, possibly due to the rather high S/N interface
resistances in this sample.\cite{BaselmansPrivate}

If we now calculate the nonequilibrium data of $I_{c}$ vs. $V_{N}$ using the
equilibrium form for $f_{L}^{0}$ in the normal reservoir, we find that the
calculation overestimates the critical current in the ``$\pi $'' state by a
large factor, and predicts too small a voltage at which the supercurrent
changes sign. This failure results from neglecting inelastic collisions
inside the wire and electron heating in the normal reservoir. \ Based on our
previous measurements of $f(E)$ in nonequilibrium mesoscopic metal wires,
\cite{PRLrelax,PierreAg} we can estimate the contributions of both inelastic
scattering and reservoir heating to the rounding of $f(E)$ in our sample.
Inelastic scattering in similar Ag wires was well described within the
framework of the Boltzmann equation using an electron-electron interaction
kernel in agreement with the theoretical form $K(E)=K_{3/2}E^{-3/2},$ but
with a prefactor $K_{3/2}\approx 0.5\,{\rm ns}^{-1}{\rm meV}^{-1/2}$, about
5 times larger than predicted by theory. Heating of the normal reservoir can
be estimated using the Wiedemann-Franz law and a simplified model of
electron-phonon scattering in the reservoir.\cite{Schonenberger,PierreAu}
The temperature of the electrons in the reservoir is given by
$T_{{\rm eff}}=\sqrt{T^{2}+b^{2}V_{N}^{2}}$\ where $b^{2}$ is proportional
to the ratio of
the reservoir sheet resistance to the wire resistance.\cite{Schonenberger}
From our sample parameters and previous measurements of similar samples,\cite
{PierreAu} we estimate $b\approx $ $1$~K/mV. Using these values of $K_{3/2}$
and $b$, we have calculated $f(E)$ and thereby $I_{c}(V_{N})$ in our sample
by solving the Boltzmann equation with the correct boundary conditions at
the S/N interfaces,\cite{PierreMAR} but neglecting proximity effect in the
bulk of the wire. The result of that calculation does not fit the data shown
in Fig.~3. A much larger value of $K_{3/2}$ $=3\,{\rm ns}^{-1}{\rm meV}%
^{-1/2}$\ provides a reasonable fit, but leaves us without a plausible
explanation for the enhanced electron-electron interactions. An alternative
approach is to use an interaction kernel of the form $K(E)=K_{2}E^{-2}$,
which describes samples containing dilute magnetic impurities.\cite
{PierreAu,Kaminski} With the value $K_{2}=0.55~{\rm ns}^{-1}$, corresponding
to a magnetic impurity concentration of about {\em 0.1~ppm}, we obtain the
solid curves shown in Fig.~3, which fit the data well at voltages up to the
crossover to the $\pi $ junction. \ Adding a reasonable $K_{3/2}$\ term to $%
K(E)$\ improves the fit only slightly at higher voltages. \ The magnetic
impurity concentration of {\em 0.1~ppm} is plausible, and will limit
$L_{\phi }$\ to about 5 $\mu $m near the Kondo temperature -- still much
larger than the distance between the two superconducting electrodes. \

The rather poor fit to the data at high voltages may reflect the fact that
the magnitude of $I_{c}$ in the $\pi $ state depends on a delicate balance
between the positive and negative parts of $j_{E}$, weighted by the precise
shape of $f(E).$ \ Fig. 4 shows $f(E)$ for $V_{N}=50$ $\mu $V, near the
maximum $\pi $ junction $I_{c}$. \ By eye $f(E)$ looks nearly like a hot
Fermi-Dirac function, but the dashed line in the figure shows that it is not.
If the sample were shorter, so that $f(E)$ maintained the staircase structure
of the dotted line in the figure, the $\pi $ junction $I_{c}$ would be much larger.

\begin{figure}[tbp]
\includegraphics[width=3in]{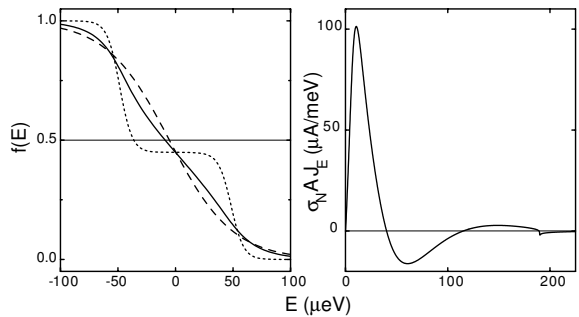}
\caption{Left: Solid line: distribution function f(E) used to
calculate the Josephson junction current in the $\pi$ state at
$V_{N} = 50 \mu V$ and T = 38 mK. Dotted line: f(E) taking into
account only reservoir heating but not energy exchange.  Dashed
line: hot Fermi-Dirac distribution. Right: Numerically calculated
$j_{E}$ (multiplied by the prefactor $\sigma _{N}A$), at the
central point of the sample, shown only for $E>0$} \label{Fig4}
\end{figure}

Fig. 4 also reveals the difference between our 3-terminal
experiment and the 4-terminal experiment of Baselmans et al. In
our sample the electrical potential is nonzero at the central
point, since the injection current flows into both S electrodes.
Hence $f(E=0)\neq \frac{1}{2}$\ \ at the central point, unlike in
Baselmans' sample. (The deviation from 1/2 is small, since the
vertical arm of our sample is much longer than the horizontal
arms.) Since the available phase space for quasiparticle energy
exchange decreases as $f(E)$ deviates from 1/2, the 3-terminal
geometry should be favorable for maximizing $I_{c}$ in the $\pi $
state. A direct measurement of this subtle effect could be made in
a 4-terminal sample. Biasing the two normal reservoirs at the same
potential $V_{N}$, rather than at asymmetric voltages $\pm V_{N}$,
would result in a current flow pattern and distribution functions
essentially equivalent to those in our 3-terminal experiment. A
comparison of the values of $I_{c}$ in the $\pi $ state under
symmetric bias $(V_{N},V_{N})$ and antisymmetric bias
$(V_{N},-V_{N})$ might reveal a subtle difference in the smearing
of f(E). We plan to explore this comparison experimentally.

We thank D. Esteve and H. Pothier for suggesting the ``dangling arm''
experiment, and I.O.~Kulik for a valuable discussion concerning electron
heating. This work was supported by NSF grants DMR-9801841 and 0104178, and
by the Keck Microfabrication Facility supported by NSF DMR-9809688.

%
%

%
%


\begin{references}
\bibitem{signofIc}  Technically it is not correct to say that $I_{c}$
changes sign. What changes sign is $I_{s}(\phi )$ for a fixed value of $\phi
$, e.g. $\phi =\pi /2$. Hence $I_{c}$\ stays positive when $V_{N}$ exceeds
40~${\rm \mu V}$, while $I_{s}$\ changes from $I_{c}\sin (\phi )$ to $%
I_{c}\sin (\phi +\pi )$.
\end{references}
\end{document}